# Adaptive Audio Watermarking via the Optimization Point of View on the Wavelet-Based Entropy

Shuo-Tsung Chen, Huang-Nan Huang[1] and Chur-Jen Chen

Department of Mathematics, Tunghai University, Taichung 40704, Taiwan

E-mail: nhuang@thu.edu.tw

*Abstract*— This study aims to present an adaptive audio watermarking method using ideas of wavelet-based entropy (WBE). The method converts low-frequency coefficients of discrete wavelet transform (DWT) into the WBE domain, followed by the calculations of mean values of each audio as well as derivation of some essential properties of WBE. A characteristic curve relating the WBE and DWT coefficients is also presented. The foundation of the embedding process lies on the approximately invariant property demonstrated from the mean of each audio and the characteristic curve. Besides, the quality of the watermarked audio is optimized. In the detecting process, the watermark can be extracted using only values of the WBE. Finally, the performance of the proposed watermarking method is analyzed in terms of signal to noise ratio, mean opinion score and robustness. Experimental results confirm that the embedded data are robust to resist the common attacks like re-sampling, MP3 compression, low-pass filtering, and amplitude-scaling.

---

[1] To whom correspondence should be addressed.

# I. INTRODUCTION

With the development of internet, illegal copying problem has become more serious and thus made digital watermarking enters as an important role to handle and deal such a critical issue. This can be seen from the fact that numerous audio and image watermarking methods have been developed recently. In audio watermarking, not only must the embedded audio have excellent quality, the hidden data should also be robust against common attacks like re-sampling, mp3 compression, low-pass filtering, amplitude scaling, etc. Under these requirements, various performances were demonstrated in either the time [1–6] or the frequency domain [6–11]. For example, in the time domain, Swanson *et al.* [2] presented a watermarking procedure where watermarks are embedded through direct modification of audio samples and Lie *et al.* [4] modified group-amplitude to achieve high robustness. It is noted that the latter technique has very low capacity since it uses three segments (1020 points) to present one bit.

In the frequency domain, Huang *et al.* [7] embedded watermarks into discrete cosine transform (DCT) coefficients and hid bar codes in the time domain as synchronization codes. Due to the limitation of embedding strength in the time domain, synchronization codes are not robust; however if synchronization codes are embedded in DCT, computational cost increases. Wu *et al.* [11] employed quantization index modulation to embed information as low-frequency sub-band coefficients of the discrete wavelet transform (DWT). This technique though improves robustness against common signal processing and noise corruption, it is yet vulnerable to amplitude and time scaling. Kim *et al.* [12] proposed a novel patchwork algorithm on the piecewise DWT constants to improve traditional patchwork algorithms.

In this work, DWT coefficients [13-15] are converted into wavelet-based entropy (WBE); the mean value of each audio and properties of WBE are calculated and derived. The relationship between WBE and DWT coefficients is also illustrated using a characteristic curve. In conclusion, the watermark can be extracted only by the mean of the WBE without the original audio. Assessment of the proposed method is based on signal-to-noise ratio (SNR), mean opinion score (MOS), embedding capacity, and bit error rate (BER). Simulation results show that not only the quality of watermarked audio is optimized, the embedded data are also robust against most signal processing and attacks.

The remaining parts of this paper are organized as follows: Section II introduces DWT and WBE, along with each audio's mean and properties of the WBE been discussed. The curve of WBE and analysis of the relationship between WBE and DWT coefficients are shown in Section III. The proposed embedding and extraction processes, as well as derivation of an optimization-based formula detecting the quality of the watermarked audio are described in Section IV. Experiments are conducted to test and confirm the proposed method. The concluding remarks are summarized in Section V.

## Ⅱ. DWT AND WBE

### A. DWT

The wavelet transform maps a function in $L^2(\mathbb{R})$ onto a scale-space plane and is obtained by the basic scaling function $\varphi_{j,m}(\cdot)$ and the wavelet basis function $\psi_{j,m}(\cdot)$ which are defined as follows:

$$\varphi_{j,m}(t) = 2^{j/2}\varphi(2^j t - m), \tag{1}$$

and

$$\psi_{j,m}(t) = 2^{j/2}\psi(2^j t - m). \tag{2}$$

A construction of two subspaces

$$V_j = \operatorname{span}\left\{\varphi_{j,m} : m \in \mathbb{Z}\right\} \tag{3}$$

and

$$W_j = \operatorname{span}\left\{\psi_{j,m} : m \in \mathbb{Z}\right\} \tag{4}$$

follows where $j$ and $m$ refer to the dilation and translation parameters. Moreover, it is a necessity that the subspaces

$$\left\{0\right\} \subset \cdots \subset V_1 \subset V_0 \subset V_{-1} \subset \cdots \subset L^2(\mathbb{R}) \tag{5}$$

form a multi-resolution analysis of $L^2(\mathbb{R})$ and the subspaces $\cdots, W_1, W_0, W_{-1}, \cdots$ stand for the orthogonal differences of the $V_m$ above. The orthogonal relations follow the existence of sequences $h = \left\{h_m\right\}_{m \in \mathbb{Z}}$ and $g = \left\{g_m\right\}_{m \in \mathbb{Z}}$ that satisfy the following identities:

$$h_m = \left\langle \varphi_{0,0}, \varphi_{-1,m} \right\rangle \text{ and } \varphi(t) = \sqrt{2}\sum_{m \in \mathbb{Z}} h_m \varphi(2t - m) \tag{6}$$

$$g_m = \left\langle \psi_{0,0}, \varphi_{-1,m} \right\rangle \text{ and } \psi(t) = \sqrt{2}\sum_{m \in \mathbb{Z}} g_m \varphi(2t - m) \tag{7}$$

where $h = \left\{h_m\right\}_{m \in \mathbb{Z}}$ and $g = \left\{g_m\right\}_{m \in \mathbb{Z}}$ denote low-pass and high-pass filters, respectively. The quadrature mirror filters (QMFs), a class of filters in digital signal processing, are selected to compute the k-level approximations [13-15].

Based on the structure mentioned above, original digital audio will be transformed into the wavelet domain with eight-level decomposition. As far as robustness, we embed synchronization codes and watermarks into lowest-frequency coefficients in level seven and eight, respectively.

### B. WBE

Suppose that the random sample $\pi_N = \{\lambda_i : 0 \leq i \leq N-1\}$ contains $N$ non-empty outcomes

each has probabilities $P(\lambda_i) = p_i, \ 0 \leq i \leq N-1$. According to information theory [16], the

measure of an uncertainty that quantifies a piece of information contained in a message is defined

as

$$M(\pi_N) = -\sum_{i=0}^{N-1} p_i \log p_i, \ 0 \leq i \leq N-1. \tag{8}$$

Similar to the definition in (8), the low-frequency DWT coefficients will be combined to identify the function defined below.

**Definition 1.** Let $C_N = \{ |c_i| : 0 \leq i \leq N-1 \}$ be a set of low-frequency DWT coefficients, the

*wavelet-based entropy* (WBE) of $C_N$ can be defined as

$$F(C_N) = -\sum_{k=0}^{N-1} z_k \log z_k \tag{9}$$

where $z_k = \dfrac{|c_k|}{\sum\limits_{k=0}^{N-1} |c_k|}, \ 0 \leq k \leq N-1,$ denotes the absolute weight of $c_k$ in $C_N$.

Since a larger $N$ results in lower embedding capacity, $N=2$ is set in the embedding process

to ensure high embedding capacity. Let $C_2 = \{ |c_0|, |c_1| \}$ be the set of DWT low-frequency

coefficients, the WBE of $C_2$ becomes

$$F(C_2) = -\left\{ \frac{|c_0|}{|c_0|+|c_1|} \log \frac{|c_0|}{|c_0|+|c_1|} + \frac{|c_1|}{|c_0|+|c_1|} \log \frac{|c_1|}{|c_0|+|c_1|} \right\}. \tag{10}$$

**III. THE INVARIANT FEATURES AND PROPERTIES OF WBE**

This section includes the invariant features and important properties of WBE. Based on the properties, a characteristic curve relates the WBE and DWT coefficients is shown.

*A. Invariant Features of WBE*

Orthogonal wavelet basis functions not only provide simple calculation in coefficients expansion but also span $L^2(\mathbb{R})$ in signal processing. As a result, audio signal $S(t) \in L^2(\mathbb{R})$ can be expressed as a series expansion of orthogonal scaling functions and wavelets. More specifically,

$$S(t) = \sum_{\ell} c_{j_0}(\ell)\varphi_{j_0,k}(t) + \sum_{k}\sum_{j=j_0}^{\infty} d_j(k)\psi_{j,k}(t) \tag{11}$$

where $c_j(\ell) = \int_{\mathbb{R}} S(t)\varphi_{j,\ell}(t)dt$ and $d_j(k) = \int_{\mathbb{R}} S(t)\psi_{j,k}(t)dt$ be the low-pass and high-pass coefficients, respectively; $j_0$ is an integer to define an interval on which $S(t)$ is piecewise constant. According to Parseval's theorem, the energy in a signal is

$$\int_{\mathbb{R}} |S(t)|^2\, dt = \sum_{\ell=-\infty}^{\infty} |c_{j_0}(\ell)|^2 + \sum_{k=-\infty}^{\infty}\sum_{j=j_0}^{\infty} |d_j(k)|^2. \tag{12}$$

It is noted that the energy in a signal can be expressed in terms of DWT coefficients [14, 15].

Let $\bar{S}(t)$ and $\tilde{S}(t)$ be the signals prior to and after amplitude scaling. Since these two signals form scale multiple of each other, we have

$$\tilde{S}(t) = \tau \cdot \bar{S}(t) \tag{13}$$

where $\tau$ being the scaling factor,

$$\tilde{S}(t) = \tau \overline{S}(t) = \tau \sum_k \overline{c}_{j_0}(k)\varphi_{j_0,k}(t) + \tau \sum_k \sum_{j=j_0}^{\infty} \overline{d}_j(k)\psi_{j,k}(t)$$

$$= \sum_k \tau \overline{c}_{j_0}(k)\varphi_{j_0,k}(t) + \sum_k \sum_{j=j_0}^{\infty} \tau \overline{d}_j(k)\psi_{j,k}(t)$$

$$= \sum_k \tilde{c}_{j_0}(k)\varphi_{j_0,k}(t) + \sum_k \sum_{j=j_0}^{\infty} \tilde{d}_j(k)\psi_{j,k}(t). \qquad (14)$$

Throughout this study, the host digital audio signal $S(n)$, $n \in \mathbb{N}$, which denotes the samples of the original audio signal $S(t)$ at the $n$-th sample time, is cut into segments on which DWT are preformed. Let $\overline{C}_2 = \{|\overline{c}_0|, |\overline{c}_1|\}$ and $\tilde{C}_2 = \{|\tilde{c}_0|, |\tilde{c}_1|\}$ be the sets of low-frequency DWT coefficients of $\overline{S}(n)$ and $\tilde{S}(n)$, which correspond to the digital audio signals prior to and after amplitude scaling attack, respectively. The relation in (14) implies that their corresponding WBEs satisfy

$$F(\tilde{C}_2) = -\left\{ \frac{|\tilde{c}_0|}{|\tilde{c}_0|+|\tilde{c}_1|} \log \frac{|\tilde{c}_0|}{|\tilde{c}_0|+|\tilde{c}_1|} + \frac{|\tilde{c}_1|}{|\tilde{c}_0|+|\tilde{c}_1|} \log \frac{|\tilde{c}_1|}{|\tilde{c}_0|+|\tilde{c}_1|} \right\}$$

$$= -\left\{ \frac{|\tau\overline{c}_0|}{|\tau\overline{c}_0|+|\tau\overline{c}_1|} \log \frac{|\tau\overline{c}_0|}{|\tau\overline{c}_0|+|\tau\overline{c}_1|} + \frac{|\tau\overline{c}_1|}{|\tau\overline{c}_0|+|\tau\overline{c}_1|} \log \frac{|\tau\overline{c}_1|}{|\tau\overline{c}_0|+|\tau\overline{c}_1|} \right\}$$

$$= -\left\{ \frac{\tau|\overline{c}_0|}{\tau|\overline{c}_0|+\tau|\overline{c}_1|} \log \frac{\tau|\overline{c}_0|}{\tau|\overline{c}_0|+\tau|\overline{c}_1|} + \frac{\tau|\overline{c}_1|}{\tau|\overline{c}_0|+\tau|\overline{c}_1|} \log \frac{\tau|\overline{c}_1|}{\tau|\overline{c}_0|+\tau|\overline{c}_1|} \right\}$$

$$= -\left\{ \frac{|\overline{c}_0|}{|\overline{c}_0|+|\overline{c}_1|} \log \frac{|\overline{c}_0|}{|\overline{c}_0|+|\overline{c}_1|} + \frac{|\overline{c}_1|}{|\overline{c}_0|+|\overline{c}_1|} \log \frac{|\overline{c}_1|}{|\overline{c}_0|+|\overline{c}_1|} \right\}$$

$$= F(\overline{C}_2). \qquad (15)$$

It is worth to mention that equation (15) indicates the nice feature of the WBE being invariant to amplitude scaling. The mean and standard deviation of $F(C_2)$, prior and after various attacks, are calculated with simulation results shown in Tables I-VII. A close look at these tables, one can conclude that the mean and standard deviation of $F(C_2)$ are approximately invariant except for the time scaling. Moreover, the low-frequency DWT coefficients in level eight have similar results to that in level seven. The novel audio watermarking proposed is based on the invariance of the mean of WBE. Since different audios have different mean values, the watermark design should depend on the mean of each audio's WBE. In other words, the proposed audio watermarking indeed demonstrates an adaptive technique.

### B. Maximum of the WBE

To find the range of the WBE, two non-negative scaling factors $\alpha_0$ and $\alpha_1$ are employed in the adjusted low-frequency DWT coefficients, namely, $\alpha C_2 = \{\alpha_0|c_0|, \alpha_1|c_1|\}$ and the adjusted WBE becomes

$$F(\alpha C_2) = -\left\{ \frac{\alpha_0|c_0|}{\alpha_0|c_0|+\alpha_1|c_1|} \log \frac{\alpha_0|c_0|}{\alpha_0|c_0|+\alpha_1|c_1|} + \frac{\alpha_1|c_1|}{\alpha_0|c_0|+\alpha_1|c_1|} \log \frac{\alpha_1|c_1|}{\alpha_0|c_0|+\alpha_1|c_1|} \right\} \quad (16)$$

or

$$F(\alpha C_2) = -\left\{ \frac{\alpha_0/\alpha_1}{\alpha_0/\alpha_1+|c_1|/|c_0|} \log(\frac{\alpha_0/\alpha_1}{\alpha_0/\alpha_1+|c_1|/|c_0|}) + \frac{|c_1|/|c_0|}{\alpha_0/\alpha_1+|c_1|/|c_0|} \log(\frac{|c_1|/|c_0|}{\alpha_0/\alpha_1+|c_1|/|c_0|}) \right\}.$$

Set $\gamma = \alpha_0/\alpha_1$ and $\mu = |c_1|/|c_0|$ to obtain

$$F(\alpha C_2) = -\left\{ \frac{\gamma}{\gamma+\mu} \log(\frac{\gamma}{\gamma+\mu}) + \frac{\mu}{\gamma+\mu} \log(\frac{\mu}{\gamma+\mu}) \right\}. \quad (17)$$

By using a change of variable,

$$x = \frac{\gamma}{\gamma+\mu} \quad (18)$$

then

$$f(x) \equiv F(\alpha C_2) = -\left\{x \log x + (1-x) \log(1-x)\right\}, \tag{19}$$

**Theorem 1.** The $f(x)$ defined in (19) has a maximum of $\log 2$, $x \in [0,1]$.

**Proof:**

1) Domain: The domain of $f(x)$ is $[0,1]$. ( Please see **APPENDIX** for detailed assertion).

2) The first derivative of $f(x)$ gives

$$f'(x) = -\left\{\log x + x \cdot \frac{1}{x} + (-1) \log(1-x) + (1-x) \cdot \frac{-1}{1-x}\right\}$$

$$= -\left\{\log x - \log(1-x)\right\} = \log \frac{1-x}{x}$$

The unique critical number, $x = \frac{1}{2}$, follows by setting $f'(x) = 0$. Since the sign of $f'(x)$ changes from positive to negative at $x = \frac{1}{2}$, $f(x)$ only increases to the left of $x = \frac{1}{2}$ and decreases to the right of $x = \frac{1}{2}$. Clearly, it follows that $f(x)$ has a unique maximum $f(\frac{1}{2}) = -\log \frac{1}{2} = \log 2$.

*C. Range of the WBE*

The nonlinear equation (19) is simply a reformulation of (16) and its characteristic curve is demonstrated in Fig. 1. According to (18), $x$ is a function of the DWT coefficients, therefore the characteristic curve also shows the relationship between the WBE and DWT coefficients. According to Theorem 1, range of $f(x)$ or $F(\alpha C_2)$ is $[0, \log 2]$ and the supremum and infimun of the WBE are $F_{\sup} = \log 2$ and $F_{\inf} = 0$.

## IV. THE PROPOSED NOVEL WATERMARKING METHOD

In the previous section, the mean of each audio and the relationship between WBE and DWT both confirm the invariant property, which serves as the kernel in designing watermarking procedure. The synchronization codes and watermarks will be embedded into the low-frequency DWT coefficients by the proposed watermarking method.

### A. Embedding Procedure

In general, watermarked audio may suffer from attacks of shifting or cropping. Hence, the watermark is arranged after converting synchronization codes into a binary pseudo-random noise (PN) sequence, $B = \{\beta_i\}$, which is embedded into the lowest frequency coefficients of its discrete wavelet transform. During the extraction procedure, the synchronization codes will be used to locate positions of embedded watermarks.

The flowchart of watermarking procedure is shown in Fig. 2 and its core, the embedding process, is described in Fig. 3. Suppose that the original audio signal $S(n)$ of length $L$ is cut into several segments on which $H$-level discrete wavelet transform is performed. The total number of lowest-frequency coefficients is $T = L / 2^H$. Let the collection of the first two lowest-frequency DWT coefficients be $C_2 = \{|c_0|, |c_1|\}$, then its WBE yields

$$F(C_2) = -\left\{ \frac{|c_0|}{|c_0| + |c_1|} \log \frac{|c_0|}{|c_0| + |c_1|} + \frac{|c_1|}{|c_0| + |c_1|} \log \frac{|c_1|}{|c_0| + |c_1|} \right\}.$$

Accordingly, the lowest-frequency DWT coefficients can be divided into $T/2$ groups and each group consists of two consecutive lowest-frequency DWT coefficients. Thus, mean value of the WBE for the original audio signal takes the value

$$F_{\text{mean}} = \left. \sum_{T/2} F(C_2) \middle/ T/2 \right. .$$ (20)

An adaptive embedding process is proposed based upon the selection key $\varepsilon$ stated as follows:

- If the embedded bit $\beta_i$ is "1", $x = x^*$ is determined using bisection method such that

$$F_{\text{mean}} + \varepsilon \leq f(x^*) \leq F_{\text{mean}} + 2\varepsilon$$ (21)

- If the embedded bit $\beta_i$ is "0", $x = x^*$ is determined using bisection method and

$$F_{\text{mean}} - 2\varepsilon \leq f(x^*) \leq F_{\text{mean}} - \varepsilon$$ (22)

where $0 < 2\varepsilon < \min\left\{F_{\text{sup}} - F_{mean}, F_{\text{mean}} - F_{\inf}\right\}$ and $f(x^*) = F(\alpha C_2)$. Since $f(x) = F(C_2)$ is continuous and monotone on $[0, 1/2]$, there always exists a real number $x^*$ such that either (21) or (22) holds for a suitable key $\varepsilon$ and makes $F_{\text{mean}} + 2\varepsilon < F_{\text{sup}}$ and $F_{\inf} < F_{\text{mean}} - 2\varepsilon$, or equivalently, $2\varepsilon < \min\{F_{\text{sup}} - F_{\text{mean}}, F_{\text{mean}} - F_{\inf}\}$. To find the numerical approximation of $x^*$, the bisection method is utilized. This approach repetitively bisects an interval and then selects the subinterval containing the root for further processing. Initially, an arbitrary small positive $\delta$ is selected to form an interval $[0 + \delta, 1/2 - \delta]$. The Intermediate Theorem states that a continuous function $f(x) - f(x^*)$ has at least one root in $[\delta, 1/2 - \delta]$ if the opposite signs condition, namely $[f(0 + \delta) - f(x^*)] \cdot [f(1/2 - \delta) - f(x^*)] < 0$, is assured. If the selected $\delta$ does not satisfy the opposite sign condition, we then increase the value of $\delta$ so to shrink the candidate interval until the condition is fulfilled. Once the candidate interval is determined, bisection method applies to solve for $x^*$. Fig. 4 shows the bisection method for solving $x^*$.

Since $\mu$ is defined to be the ratio of $|c_1|$ to $|c_0|$, $\gamma$ can also be easily found by

$$\gamma = -\frac{\mu x^*}{x^* - 1},\tag{23}$$

and have either $F_{\text{mean}} + \varepsilon \leq F(\alpha C_2) \leq F_{\text{mean}} + 2\varepsilon$ or $F_{\text{mean}} - 2\varepsilon \leq F(\alpha C_2) \leq F_{\text{mean}} - \varepsilon$

satisfied. In the later section, discussions on values of $\alpha_0$ and $\alpha_1$ will be addressed so they not only maximize the SNR but also keep the ratio $\gamma$ unchanged. Once the values of $\alpha_0$ and $\alpha_1$ are computed, the DWT coefficients $C_2 = \{|c_0|, |c_1|\}$ is modified as $\overline{C}_2 = \{|\overline{c}_0|, |\overline{c}_1|\} = \{\alpha_0 |c_0|, \alpha_1 |c_1|\}$.

The embedding process consists of the following steps:

Step 1. Set $\delta = 0.001$ and let $[\delta, 1/2 - \delta]$ be the initial candidate interval to seek for

approximate $x^*$ via the bisection method. An initial estimate of $x^*$ is set to be

$(\delta + 1/2 - \delta)/2$.

Step 2. Check whether the following conditions satisfy.

If $\beta_i = 1$, then $F_{\text{mean}} + \varepsilon \leq f(x^*) \leq F_{\text{mean}} + 2\varepsilon$.

If $\beta_i = 0$, then $F_{\text{mean}} - 2\varepsilon \leq f(x^*) \leq F_{\text{mean}} - \varepsilon$.

If either one of the condition holds, the embedding process is finished, otherwise,

continue to Step3.

Step 3. Use the bisection method to update $x^*$ and then go to Step 2.

Lastly, from the set of modified DWT coefficients $\overline{C}_2 = \{|\overline{c}_0|, |\overline{c}_1|\}$, one can apply the inverse discrete wavelet transform (IDWT) to generate the time-domain watermarked audio $\overline{S}(n)$.

## B. Extraction Procedure

Fig. 5 demonstrates the watermark extraction procedure. Firstly, one repeatedly applies the DWT together with the extraction procedure to a watermarked signal, say $\bar{S}(n)$, to obtain the synchronization codes. In this step, every two consecutive DWT coefficients will be put together as a set $\bar{C}_2 = \left\{ |\bar{c}_0|, |\bar{c}_1| \right\}$ without ever sorting lowest frequency. To extract the watermark, Definition 1 with $N = 2$ is utilized as follows.

- If $F_{\text{mean}} + \varepsilon \leq F(\bar{C}_2) \leq F_{\text{mean}} + 2\varepsilon$, the extracted value $\beta_i = 1$.

- If $F_{\text{mean}} - 2\varepsilon \leq F(\bar{C}_2) \leq F_{\text{mean}} - \varepsilon$, the extracted value $\beta_i = 0$.

where $\bar{C}_2$ denotes the embedded coefficients of $C_2$ and $\varepsilon$ is the selected key in the embedding procedure. After finding the synchronization codes, repetition of extraction process carries over to detect the watermark.

## C. Optimization for Scaling Factors

In general, the quality of the watermarked audio can be determined using SNR as a performance index formulated as

$$\text{SNR} = 10 \log \left( \frac{\left\| S(n) \right\|_2^2}{\left\| \bar{S}(n) - S(n) \right\|_2^2} \right) \tag{24}$$

or equivalently,

$$\text{SNR} = -10 \log \left( \frac{\left\| \bar{S}(n) - S(n) \right\|_2^2}{\left\| S(n) \right\|_2^2} \right) \tag{25}$$

where $S(n)$ and $\bar{S}(n)$ denote the original and the watermarked audio signal, respectively. To simplify the optimization operations later, equation (25) is selected to be the performance index instead of using the SNR directly. Since the orthogonal wavelets are applied in the embedding process, the corresponding performance index in wavelet domain can be expressed as:

$$-10\log\left\{\frac{\sum_{i=0}^{N-1}\left(|\,\overline{c}_i\,|-|\,c_i\,|\right)^2}{\sum_{i=0}^{N-1}|c_i|^2}\right\} \qquad (26)$$

where $c_i$ and $\overline{c}_i$ are the corresponding DWT coefficients. In this study, each binary bit is embedded using two consecutive DWT coefficients, i.e., $N=2$, the performance index in (26) equates to

$$-10\log\left\{\frac{(|\overline{c}_0|-|c_0|)^2+(|\overline{c}_1|-|c_1|)^2}{|c_0|^2+|c_1|^2}\right\}$$

or

$$-10\log\left\{\frac{(\alpha_0-1)^2|c_0|^2+(\alpha_1-1)^2|c_1|^2}{|c_0|^2+|c_1|^2}\right\} \qquad (27)$$

with the selection of $|\,\overline{c}_0\,|=\alpha_0\,|\,c_0\,|$ and $|\,\overline{c}_1\,|=\alpha_1\,|\,c_1\,|$.

In order to have the best audio quality, we must maximize the SNR which is equivalent to minimize

$$\frac{(\alpha_0-1)^2|c_0|^2+(\alpha_1-1)^2|c_1|^2}{|c_0|^2+|c_1|^2}$$

due to the fact that the log function is one-to-one. Since the ratio $\gamma=\alpha_0\,/\,\alpha_1$ is pre-determined, the optimization problem arises:

$$\textit{minimize} \qquad \frac{(\alpha_0-1)^2|c_0|^2+(\alpha_1-1)^2|c_1|^2}{|c_0|^2+|c_1|^2} \qquad (28a)$$

$$\textit{subjected to} \qquad \alpha_0 - \gamma\alpha_1 = 0 \qquad\qquad (28b)$$

Let $\lambda$ denote the Lagrange multiplier, the optimization problem (28) is converted to finding the minimum of the unconstrained function

$$I(\alpha_0, \alpha_1, \lambda) = \frac{(\alpha_0 - 1)^2 |c_0|^2 + (\alpha_1 - 1)^2 |c_1|^2}{|c_0|^2 + |c_1|^2} + \lambda(\alpha_0 - \gamma\alpha_1)$$

or equivalent to finding the minimum of

$$J(\alpha_0, \alpha_1, \lambda) = (\alpha_0 - 1)^2 |c_0|^2 + (\alpha_1 - 1)^2 |c_1|^2 + \lambda(\alpha_0 - \gamma\alpha_1)(|c_0|^2 + |c_1|^2) \qquad (29)$$

since $|c_0|^2 + |c_1|^2$ is known in advance. According to the Lagrange Principle, necessary conditions for existence of minimum of $J(\alpha_0, \alpha_1, \lambda)$ are

$$\frac{\partial J}{\partial \alpha_0} = 2|c_0|^2 (\alpha_0 - 1) + \lambda(|c_0|^2 + |c_1|^2) = 0$$

$$\frac{\partial J}{\partial \alpha_1} = 2|c_1|^2 (\alpha_1 - 1) - \gamma\lambda(|c_0|^2 + |c_1|^2) = 0$$

$$\frac{\partial J}{\partial \lambda} = (\alpha_0 - \gamma\alpha_1)(|c_0|^2 + |c_1|^2) = 0$$

which leads to the optimal scaling values $\alpha_0^*, \alpha_1^*$:

$$\alpha_0^* = \gamma\left(\frac{\gamma|c_0|^2 + |c_1|^2}{\gamma^2|c_0|^2 + |c_1|^2}\right), \quad \alpha_1^* = \frac{\gamma|c_0|^2 + |c_1|^2}{\gamma^2|c_0|^2 + |c_1|^2} \qquad (30)$$

with the ratio $\gamma$ given in (23). Direct verification shows that these optimal scaling values do provide minimum of $J(\alpha_0, \alpha_1, \lambda)$. Consequently, one achieves the maximal value of (26). In other words, the maximal value of the SNR in (24) is achieved with the DWT coefficients of $\bar{S}(n)$ selected as $\bar{C}_2 = \{|\bar{c}_0|, |\bar{c}_1|\} = \{\alpha_0^*|c_0|, \alpha_1^*|c_1|\}$.

## V. EXPERIMENTAL RESULTS AND DISCUSSIONS

This section investigates the performance of the proposed audio watermarking technique by the SNR, MOS, embedding capacity, and BER. The quality of watermarked audio is mathematically measured by the SNR defined in (24). To test the watermarked audio quality in practice, both original and watermarked audio were provided to ten listeners who score each audio by MOS values, as shown in Table VIII.

For the time domain technique, the method proposed by Lie *et al.* [4] provides strong robustness under re-sampling and amplitude scaling attacks due to their embedding on low-frequency amplitude modification. The quantization-based method proposed by Wu *et al.* [11] is an adaptive watermarking technique that has shown good watermarked audio quality and strong robustness comparing to other approaches. This explains why we mainly compare our technique with these two papers for time- and DWT-domain techniques, respectively.

In this study, we use four kinds of music (popular, symphony, piano, and dance) that has 16-bit mono audio sampled at 44.1 kHz. Each audio is cut into four non-overlapping segments and the threshold (i.e., the secret key) $\varepsilon$ is set to be 0.03. Table IX summarizes the difference in DWT level, the SNR, MOS and embedding capacity among the methods proposed by Lie *et al.* [4], Wu. *et al.* [11] and the proposed work. To achieve high robustness, Lie *et al.* [4] uses three segments (1020 points) to present one bit in the time domain. Since the number of lowest-frequency coefficients in level seven doubles, the SNR in level seven is smaller than that in level eight. However, embedding capacity in level seven doubles that in level eight.

Finally, BER is introduced to measure the robustness:

$$\text{BER} = \frac{B_{\text{error}}}{B_{\text{total}}} \times 100\% \; ,$$

where $B_{\text{error}}$ and $B_{\text{total}}$ denote the number of error bits and the number of total bits, respectively. To test the resistance of common attacks, some experiments are conducted. The five types of attacks, such as re-sampling, MP3 compression, low-pass filtering, amplitude scaling, and time scaling, will be introduced in the next paragraph.

(1) *Re-sampling:* In Table X, we dropped the sampling rate of the watermarked audio from 44.1 kHz to 22.05 kHz and then rose back to 44.1 kHz by interpolation. Similarly, the sampling rate varies from 44.1 kHz to 11.025 kHz, and 8 kHz, and then back to 44.1 kHz. It is noted that the BER in level eight and seven increases from 3.4% to 15.7%.

(2) *MP3 compression:* MP3 compression is the most popular technique for audio compression. In Table XI, we apply MP3 compression with different bit rates to the watermarked audio and the results in level eight is slightly better than that in level seven.

(3) *Low-pass filtering:* Table XII shows the effect of low-pass filter with cutoff frequency 3 kHz. The result in level eight is similar to that in level seven. In comparison, our method has similar robustness to Lie's and Wu's work in [4] and [11].

(4) *Amplitude scaling:* Since a large scaling factor results in saturation, we set the scaling factor

$\tau$ in (13) as 0.2, 0.8, 1.1, and 1.2. The experimental results in Table XIII shows the proposed method provides strong robustness. Although the simple content of piano enables the error rate of method [11] to decline rapidly, it is still weak than ours.

(5) *Time scaling:* The watermarked audios are scaled by -5%, -2%, 2%, and 5%. Table XIV shows that BER is approximately to be 40% due to the reason that the each audio's WBE mean vary irregularly. However, the results of our method are better than method [11].

## VI. CONCLUSION

In this paper, a novel audio watermarking technique is proposed where the information is embedded by each audio's WBE mean. For the watermarking design, properties of WBE and the characteristic curve between WBE and DWT coefficients were addressed. Moreover, the proposed method does not need the original audio for watermark detection. The performance of proposed method is assessed by the SNR, MOS, embedding capacity, and BER. All experimental results show that the embedded data are robust against most attacks.

## APPENDIX

In this appendix, we mainly focus on the domain of

$$f(x) = -\left\{ x \log x + (1-x)\log(1-x) \right\}.$$

Obviously, $f(x)$ is only meaningful on $(0,1)$ since the domain of $\log x$ is $(0,\infty)$. To include both endpoints in the domain, the following limits need to be carried out.

$$\lim_{x \to 0^+} f(x) = -\lim_{x \to 0^+} \left\{ x \log x + (1-x) \log(1-x) \right\}$$

$$= \lim_{x \to 0^+} \left\{ \frac{\log x}{1/x} \right\} + 1 \cdot \log 1 = \lim_{x \to 0^+} \left\{ \frac{1/x}{-1/x^2} \right\} = \lim_{x \to 0^+} \left\{ -\frac{x^2}{x} \right\} = 0$$

Similarly,

$$\lim_{x \to 1^-} f(x) = -\lim_{x \to 1^-} \left\{ x \log x + (1-x) \log(1-x) \right\} = 0.$$

Define the function $f(x)$ as

$$f(x) = \begin{cases} -\left\{ x \log x + (1-x) \log(1-x) \right\}, & x \in (0,1), \\ 0, & x = 0 \text{ or } x = 1, \end{cases}$$

then the domain of the continuous function $f(x)$ can be extended to be $[0,1]$.

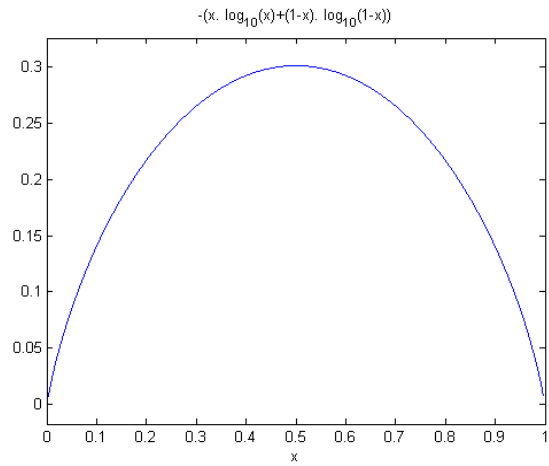

Fig. 1. The relationship between the WBE and DWT coefficients.

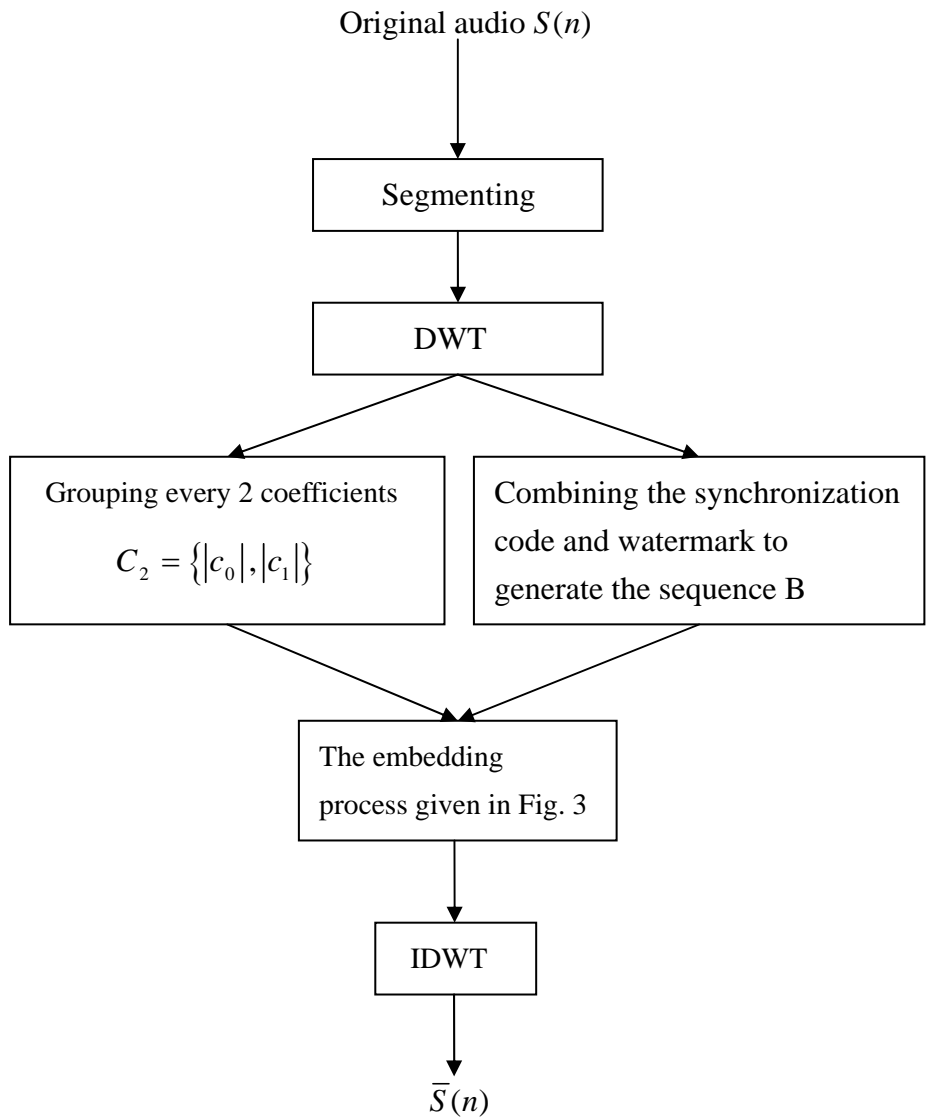

Fig. 2. The watermark embedding procedure.

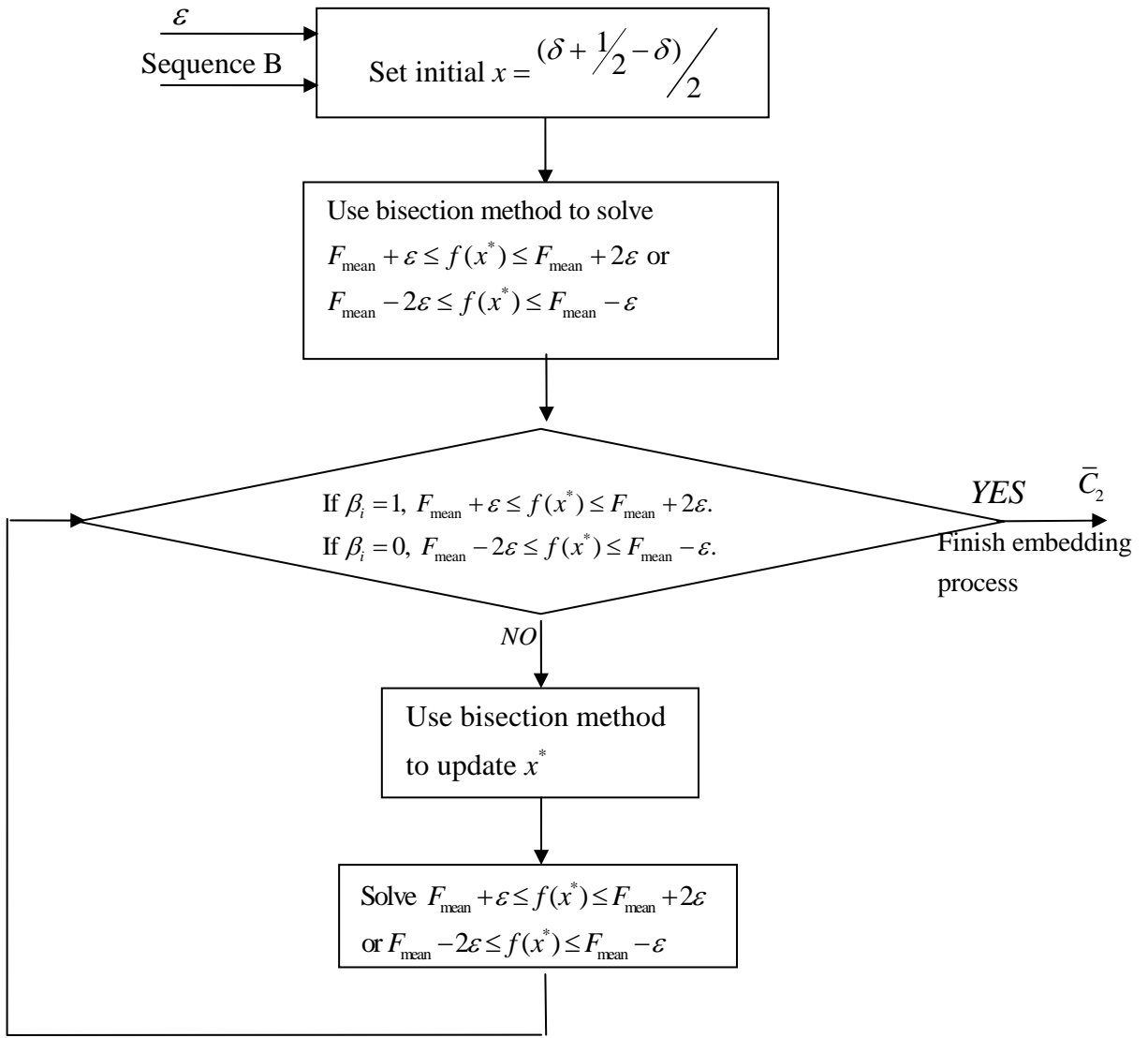

Fig. 3. The embedding process.

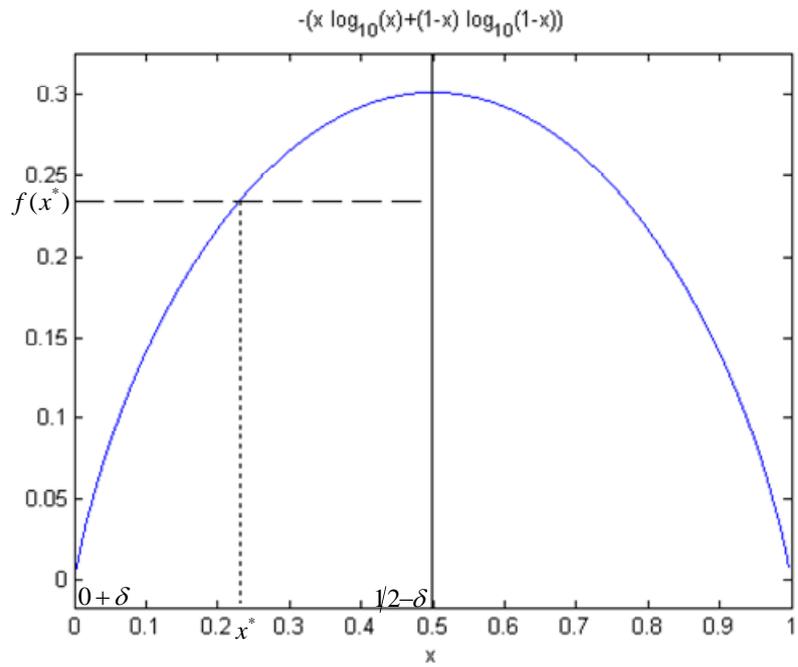

Fig. 4. The bisection method for solving $x^*$.

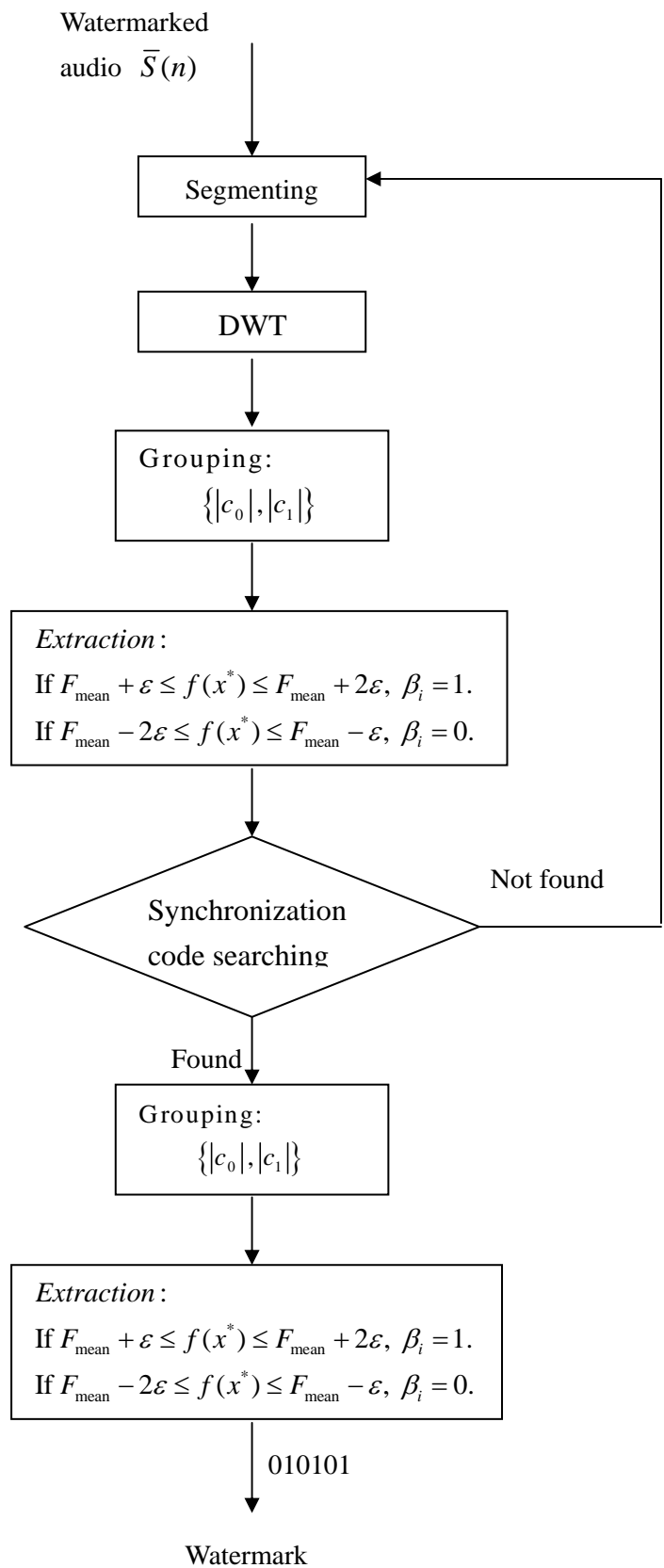

Fig. 5. The watermark extracting procedure.

**TABLE I**

THE WBE MEAN AND STANDARD DEVIATION FOR THREE AUDIOS IN DWT LEVEL 7

| Audio type | Mean of WBE | Standard deviation of WBE |
|---|---|---|
| Popular | 0.2547 | 0.0651 |
| Symphony | 0.2353 | 0.0736 |
| Piano | 0.2472 | 0.0048 |

**TABLE II**

THE WBE MEAN OF POPULAR FOR VARIOUS ATTACKS IN DWT LEVEL 7

| Re-sampling | | MP3 compression | | Low-pass filter | | Amplitude scaling | | Time scaling | |
|---|---|---|---|---|---|---|---|---|---|
| Rate (Hz) | Mean | Bit rate (kbps) | Mean | Cut-off (kHz) | Mean | Scaling factor | Mean | Scaling factor | Mean |
| 22050 | 0.2543 | 128 | 0.2547 | 3 | 0.2558 | 0.2 | 0.2547 | -5 | 0.2548 |
| 11025 | 0.2548 | 112 | 0.2547 | 5 | 0.2558 | 0.8 | 0.2547 | -2 | 0.2540 |
| 8000 | 0.2548 | 96 | 0.2548 | | | 1.1 | 0.2547 | 2 | 0.2537 |
| | | 80 | 0.2548 | | | 1.2 | 0.2547 | 5 | 0.2552 |

**TABLE III**

THE WBE MEAN OF SYMPHONY FOR VARIOUS ATTACKS IN DWT LEVEL 7

| Re-sampling | | MP3 compression | | Low-pass filter | | Amplitude scaling | | Time scaling | |
|---|---|---|---|---|---|---|---|---|---|
| Rate (Hz) | Mean | Bit rate (kbps) | Mean | Cut-off (kHz) | Mean | Scaling factor | Mean | Scaling factor | Mean |
| 22050 | 0.2352 | 128 | 0.2352 | 3 | 0.2353 | 0.2 | 0.2353 | -5 | 0.2357 |
| 11025 | 0.2350 | 112 | 0.2352 | 5 | 0.2354 | 0.8 | 0.2353 | -2 | 0.2389 |
| 8000 | 0.2350 | 96 | 0.2351 | | | 1.1 | 0.2353 | 2 | 0.2334 |
| | | 80 | 0.2351 | | | 1.2 | 0.2353 | 5 | 0.2377 |

**TABLE IV**

THE WBE MEAN OF PIANO FOR VARIOUS ATTACKS IN DWT LEVEL 7

| Re-sampling | | MP3 compression | | Low-pass filter | | Amplitude scaling | | Time scaling | |
|---|---|---|---|---|---|---|---|---|---|
| Rate (Hz) | Mean | Bit rate (kbps) | Mean | Cut-off (kHz) | Mean | Scaling factor | Mean | Scaling factor | Mean |
| 22050 | 0.2483 | 128 | 0.2481 | 3 | 0.2485 | 0.2 | 0.2472 | -5 | 0.2494 |
| 11025 | 0.2488 | 112 | 0.2482 | 5 | 0.2508 | 0.8 | 0.2472 | -2 | 0.2495 |
| 8000 | 0.2488 | 96 | 0.2482 | | | 1.1 | 0.2472 | 2 | 0.2480 |
| | | 80 | 0.2482 | | | 1.2 | 0.2472 | 5 | 0.2483 |

**TABLE V**

THE WBE STANDARD DEVIATION OF POPULAR FOR VARIOUS ATTACKS
IN DWT LEVEL 7

| Re-sampling | | MP3 compression | | Low-pass filter | | Amplitude scaling | | Time scaling | |
|---|---|---|---|---|---|---|---|---|---|
| Rate (Hz) | Standard deviation | Bit rate (kbps) | Standard deviation | Cut-off (kHz) | Standard deviation | Scaling factor | Standard deviation | Scaling factor | Standard deviation |
| 22050 | 0.0653 | 128 | 0.0650 | 3 | 0.0652 | 0.2 | 0.0651 | -5 | 0.0657 |
| 11025 | 0.0648 | 112 | 0.0649 | 5 | 0.0653 | 0.8 | 0.0651 | -2 | 0.0662 |
| 8000 | 0.0648 | 96 | 0.0649 | | | 1.1 | 0.0651 | 2 | 0.0663 |
| | | | | | | 1.2 | 0.0651 | 5 | 0.0669 |

**TABLE VI**

THE WBE STANDARD DEVIATION OF SYMPHONY FOR VARIOUS ATTACKS
IN DWT LEVEL 7

| Re-sampling | | MP3 compression | | Low-pass filter | | Amplitude scaling | | Time scaling | |
|---|---|---|---|---|---|---|---|---|---|
| Rate (Hz) | Standard deviation | Bit rate (kbps) | Standard deviation | Cut-off (kHz) | Standard deviation | Scaling factor | Standard deviation | Scaling factor | Standard deviation |
| 22050 | 0.0740 | 128 | 0.0738 | 3 | 0.0736 | 0.2 | 0.0736 | -5 | 0.0730 |
| 11025 | 0.0741 | 112 | 0.0739 | 5 | 0.0736 | 0.8 | 0.0736 | -2 | 0.0721 |
| 8000 | 0.0741 | 96 | 0.0739 | | | 1.1 | 0.0736 | 2 | 0.0750 |
| | | | | | | 1.2 | 0.0736 | 5 | 0.0752 |

### TABLE VII
### THE WBE STANDARD DEVIATION OF PIANO FOR VARIOUS ATTACKS
### IN DWT LEVEL 7

| Re-sampling | | MP3 compression | | Low-pass filter | | Amplitude scaling | | Time scaling | |
|---|---|---|---|---|---|---|---|---|---|
| Rate (Hz) | Standard deviation | Bit rate (kbps) | Standard Deviation | Cut-off (kHz) | Standard deviation | Scaling factor | Standard deviation | Scaling factor | Standard deviation |
| 22050 | 0.0047 | 128 | 0.0046 | 3 | 0.0050 | 0.2 | 0.0048 | -5 | 0.0047 |
| 11025 | 0.0046 | 112 | 0.0047 | 5 | 0.0045 | 0.8 | 0.0048 | -2 | 0.0045 |
| 8000 | 0.0046 | 96 | 0.0047 | | | 1.1 | 0.0048 | 2 | 0.0046 |
| | | | | | | 1.2 | 0.0048 | 5 | 0.0050 |

### TABLE VIII
### MEAN OPINION SCORE (MOS)

| MOS | Quality | Impairment |
|---|---|---|
| 5 | Excellent | Imperceptible |
| 4 | Good | Perceptible but not annoying |
| 3 | Fair | Slightly annoying |
| 2 | Poor | Annoying |
| 1 | Bad | Very annoying |

### TABLE IX
### DOMAIN, SNR, MOS, AND EMBEDDING CAPACITY

| | Reference paper[4] | Reference paper[11] | Proposed Method | Proposed Method |
|---|---|---|---|---|
| Domain | Time domain | DWT level - 8 | DWT level 7 | DWT level 8 |
| SNR(dB) | 24.5(popular) 22.2(symphony) 19.2(piano) 19.6(dance) | 24.3(popular) 24.9(symphony) 26.3(piano) 36.3(dance) | 22.3(popular) 20.2(symphony) 18.3(piano) 11.9(dance) | 34.4(popular) 21.7(symphony) 36.4(piano) 13.5(dance) |
| MOS | 4.4(popular) 4.4(symphony) 4.1(piano) 4.1(dance) | 4.4(popular) 4.6(symphony) 4.4(piano) 5.0(dance) | 4.4(popular) 4.2(symphony) 4.4(piano) 4.0(dance) | 4.9(popular) 4.2(symphony) 4.9(piano) 4.0(dance) |
| Embedding capacity | 500bits/ 11.609seconds | 2000bits/ 11.609seconds | 2000bits/ 11.609seconds | 1000bits/ 11.609seconds |

**TABLE X**

BER (%) FOR RE-SAMPLING

| Re-sampling Rate(Hz) | Audio type | Reference paper[4] | Reference paper[11] | Proposed Method | |
|---|---|---|---|---|---|
| | | | | DWT level 7 | DWT level 8 |
| 22050 | popular | 0.3 | 0.8 | 9.1 | 8.7 |
| | symphony | 1.3 | 1.5 | 3.6 | 3.4 |
| | piano | 0.4 | 0.4 | 8.2 | 8.1 |
| | dance | 2.0 | 12.6 | 4.9 | 4.8 |
| 11025 | popular | 0.3 | 2.3 | 14.9 | 14.2 |
| | symphony | 2.4 | 3.2 | 7.5 | 7.1 |
| | piano | 0.7 | 1.2 | 12.7 | 11.8 |
| | dance | 7.9 | 28.3 | 9.9 | 9.6 |
| 8000 | popular | 0.4 | 3.6 | 15.7 | 15.2 |
| | symphony | 3.1 | 4.3 | 7.3 | 6.9 |
| | piano | 0.9 | 2.8 | 15.6 | 14.9 |
| | dance | 11.0 | 27.5 | 9.4 | 9.4 |

**TABLE XI**

BER (%) FOR MP3 COMPRESSION

| Bit Rate (kbps) | Audio type | Reference paper[4] | Reference paper[11] | Proposed Method | |
|---|---|---|---|---|---|
| | | | | DWT level 7 | DWT level 8 |
| 128 | popular | 0.3 | 0.4 | 3.2 | 3.0 |
| | symphony | 0.3 | 0.4 | 3.3 | 3.0 |
| | piano | 1.1 | 3.4 | 6.7 | 6.6 |
| | dance | 2.8 | 7.2 | 6.7 | 6.7 |
| 112 | popular | 0.3 | 1.6 | 2.3 | 2.2 |
| | symphony | 1.6 | 2.3 | 4.6 | 4.3 |
| | piano | 2.0 | 4.3 | 8.4 | 8.2 |
| | dance | 3.1 | 9.6 | 7.9 | 7.8 |
| 96 | popular | 1.6 | 2.7 | 3.5 | 3.3 |
| | symphony | 2.9 | 3.3 | 6.9 | 6.6 |
| | piano | 2.1 | 4.9 | 9.2 | 9.0 |
| | dance | 3.2 | 9.6 | 7.8 | 7.8 |

**TABLE XII**

BER (%) FOR LOW-PASS FILTERING

| Cut-off frequency | Audio type | Reference paper[4] | Reference paper[11] | Proposed Method | |
|---|---|---|---|---|---|
| | | | | DWT level 7 | DWT level 8 |
| 3 | popular | 19.9 | 22.4 | 26.7 | 26.6 |
| | symphony | 21.2 | 23.2 | 27.8 | 28.2 |
| | piano | 26.3 | 25.9 | 31.4 | 29.2 |
| | dance | 20.5 | 34.7 | 27.1 | 26.4 |

**TABLE XIII**

BER (%) FOR AMPLITUDE SCALING

| Scaling factor($\alpha$) | Audio type | Reference paper[4] | Reference paper[11] | Proposed Method | |
|---|---|---|---|---|---|
| | | | | DWT level 7 | DWT level 8 |
| 0.2 | popular | 0.6 | 51.3 | 0.4 | 0.3 |
| | symphony | 0.9 | 41.7 | 0.7 | 0.6 |
| | piano | 0.7 | 49.3 | 0.4 | 0.3 |
| | dance | 0.4 | 36.8 | 0.2 | 0.1 |
| 0.8 | popular | 0.5 | 4.1 | 0.4 | 0.4 |
| | symphony | 0.7 | 39.2 | 0.6 | 0.6 |
| | piano | 0.6 | 1.2 | 0.3 | 0.3 |
| | dance | 0.2 | 31.4 | 0.1 | 0.1 |
| 1.1 | popular | 0.4 | 0.6 | 0.3 | 0.3 |
| | symphony | 0.6 | 24.1 | 0.6 | 0.5 |
| | piano | 0.5 | 0.3 | 0.3 | 0.2 |
| | dance | 0.1 | 36.4 | 0.1 | 0.05 |
| 1.2 | popular | 0.5 | 2.4 | 0.4 | 0.4 |
| | symphony | 0.6 | 31.7 | 0.6 | 0.5 |
| | piano | 0.6 | 0.5 | 0.3 | 0.2 |
| | dance | 0.2 | 38.1 | 0.1 | 0.05 |

**TABLE XIV**

BER (%) FOR TIME SCALING

| Time scaling (%) | Audio type | Reference paper[4] | Reference paper[11] | Proposed Method | |
|---|---|---|---|---|---|
| | | | | DWT level 7 | DWT level 8 |
| -5 | popular | 38.9 | 52.4 | 42.2 | 41.1 |
| | symphony | 38.9 | 51.2 | 40.1 | 39.6 |
| | piano | 42.6 | 51.1 | 41.8 | 41.2 |
| | dance | 47.3 | 50.3 | 46.8 | 46.9 |
| -2 | popular | 37.6 | 49.3 | 41.9 | 41.3 |
| | symphony | 38.5 | 51.1 | 40.7 | 40.2 |
| | piano | 41.8 | 49.3 | 41.2 | 40.5 |
| | dance | 47.2 | 50.2 | 46.7 | 46.6 |
| 2 | popular | 38.7 | 50.0 | 39.9 | 39.6 |
| | symphony | 38.8 | 54.4 | 40.4 | 38.4 |
| | piano | 40.5 | 48.7 | 40.9 | 40.2 |
| | dance | 47.2 | 52.1 | 46.3 | 46.2 |
| 5 | popular | 38.8 | 52.3 | 40.1 | 39.8 |
| | symphony | 39.8 | 51.6 | 40.6 | 38.9 |
| | piano | 42.1 | 50.7 | 42.3 | 41.6 |
| | dance | 48.6 | 53.2 | 48.0 | 47.8 |